\documentclass{ws-ijmpe}
\usepackage{epsf}

\begin{document}

\markboth{U.~Baur}{Precision Calculations for Future Colliders}

\catchline{}{}{}{}{}

\title{PRECISION CALCULATIONS FOR FUTURE COLLIDERS}

\author{\footnotesize U.~Baur}

\address{Physics Department, State University of New York at Buffalo,
Buffalo, NY 14260, USA\\
baur@buffalo.edu}

\maketitle

\begin{history}
UB-HET-07-01\\
January~2007
\end{history}

\begin{abstract}
I discuss the motivations for, and the status of, precision calculations
for the Large Hadron Collider (LHC) and the planned International Linear
Collider (ILC). 
\end{abstract}

\section{Why do we care?}

In less than a year, the CERN Large Hadron Collider (LHC) will begin
operation. The 
LHC will collide protons at a center-of-mass energy of $\sqrt{s}=14$~TeV
with a design luminosity of ${\cal L}=10^{34}\,{\rm cm^{-2}\,
s^{-1}}$. This represents an increase of a factor of seven in energy and
a factor of 100 in luminosity over the Fermilab Tevatron.  
With its unprecedented energy and luminosity, the LHC promises to
revolutionize particle physics.  It will unveil the mechanism of
electroweak symmetry breaking (EWSB) and shed light on the physical
processes that are responsible for the origin of mass.  The LHC holds
the potential to make dark matter in the laboratory and perhaps even to
reveal extra dimensions of space.  Its reach for uncovering new
phenomena is dramatically higher than that of all previous
accelerators. The LHC truly will be a discovery machine. 

For the next decade, the particle physics community is planning to
build a linear $e^+e^-$ collider with a center of mass
energy in the range of $500-1000$~GeV and a luminosity of ${\cal
L}=2\times 10^{34}\,{\rm cm^{-2}\, s^{-1}}$. An $e^+e^-$ collider will
provide a cleaner environment than a hadron collider and will complement
the LHC in its search for new physics\cite{Aguilar-Saavedra:2001rg}.

\begin{figure}[t]
\epsfxsize=12.cm
\epsfclipon
\centerline{
\epsffile[50 200 550 600]{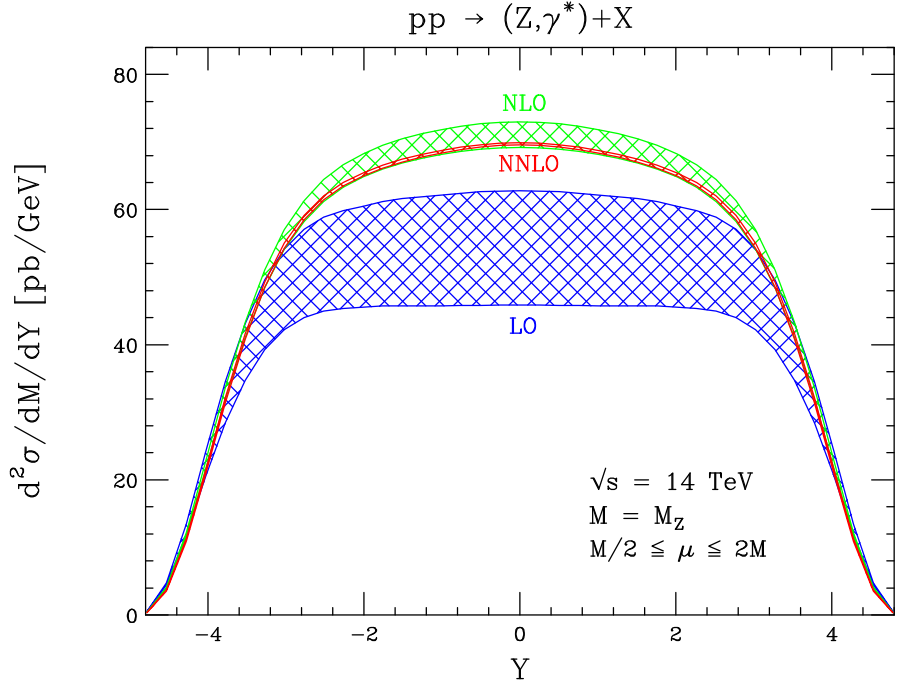}
}
\epsfclipoff
\vspace*{8pt}
\caption{\label{fig:one}The $Z$ boson rapidity distribution at the
LHC. Shown are the LO, NLO and NNLO predictions (from Ref.~[2]).}
\end{figure}

To uncover the mechanism of EWSB and discover new physics at the LHC, it is
necessary to have accurate theoretical calculations of Standard Model
(SM) processes and
new physics signatures. The final states of many processes are quite
complex at the LHC. The lowest-order (LO) predictions for many SM processes
exhibit a significant dependence on the unphysical
renormalization and factorization scales that can be traced to the
truncation of the perturbation series.  The scale dependence can be
reduced by calculating observables to higher order in 
perturbation theory.  Higher-order QCD and, in some cases, electroweak (EW)
radiative corrections are needed for accurate SM predictions. Sometimes,
such as for $W$ and $Z$
production\cite{Anastasiou:2003ds}, the reduction of the scale
dependence once higher order QCD 
corrections are taken into account is dramatic. This is illustrated in
Fig.~\ref{fig:one} where the $Z$ boson rapidity distribution at the LHC is
shown. While the LO cross section varies by about $50\%$ for a 
renormalization/factorization scale $\mu$ in the range $M_Z/2\leq\mu\leq
2M_Z$, the uncertainty at next-to-leading order (NLO) is reduced to
about 10\%, and at next-to-next-to-leading order (NNLO) to
about 1\%.

Although much has been accomplished in recent years, much remains to be
done in order to ensure that the full physics potential of the LHC can
be utilized. Recent results relevant for the LHC are discussed in
Sec.~3.1. Here, without going into any details, I give a time ordered ``LHC
shopping list'' of precision calculations which are still needed:

\begin{enumerate}

\item For $10-30$~fb$^{-1}$ (2009 -- 2010):

\begin{enumerate}

\item compute full NLO QCD corrections to $pp\to t\bar t\to b\bar b+4f$

\item compute full tree level calculation of $t\bar tWjj$ production

\item compute NLO QCD corrections to $t\bar tj$, $t\bar t\gamma$, $t\bar tb\bar
b$, $t\bar tjj$ and $WWjj$ production

\item resum QCD corrections to $qq'\to qq'H$

\end{enumerate}

\item For 300~fb$^{-1}$ (2012 -- 2013): compute NLO QCD corrections to $gg\to
HH$, $t\bar tW$ and $t\bar tZ$ production 

\item For 3000~fb$^{-1}$ (SuperLHC, $>2015$): compute NLO QCD
corrections to $WWWjj$, 
$jj\gamma\gamma$ and $Q\bar Q\gamma j$ production

\end{enumerate}

The enormous center of mass energy of the LHC makes it an ideal tool to
search for new particles which are a common prediction of all new
physics scenarios. On the other hand, 
the cleaner environment of the ILC will make it easier to
precisely measure SM observables, such as the $W$ mass. Their
measurement is expected to yield 
complementary information on new physics. New heavy particles, which are
a common prediction of all beyond-the-SM models, generally contribute to
observables via virtual radiative corrections, and thus lead to
small deviations from the SM predictions. 

\begin{figure}[t]
\epsfxsize=12.cm
\epsfclipon
\centerline{
\epsffile{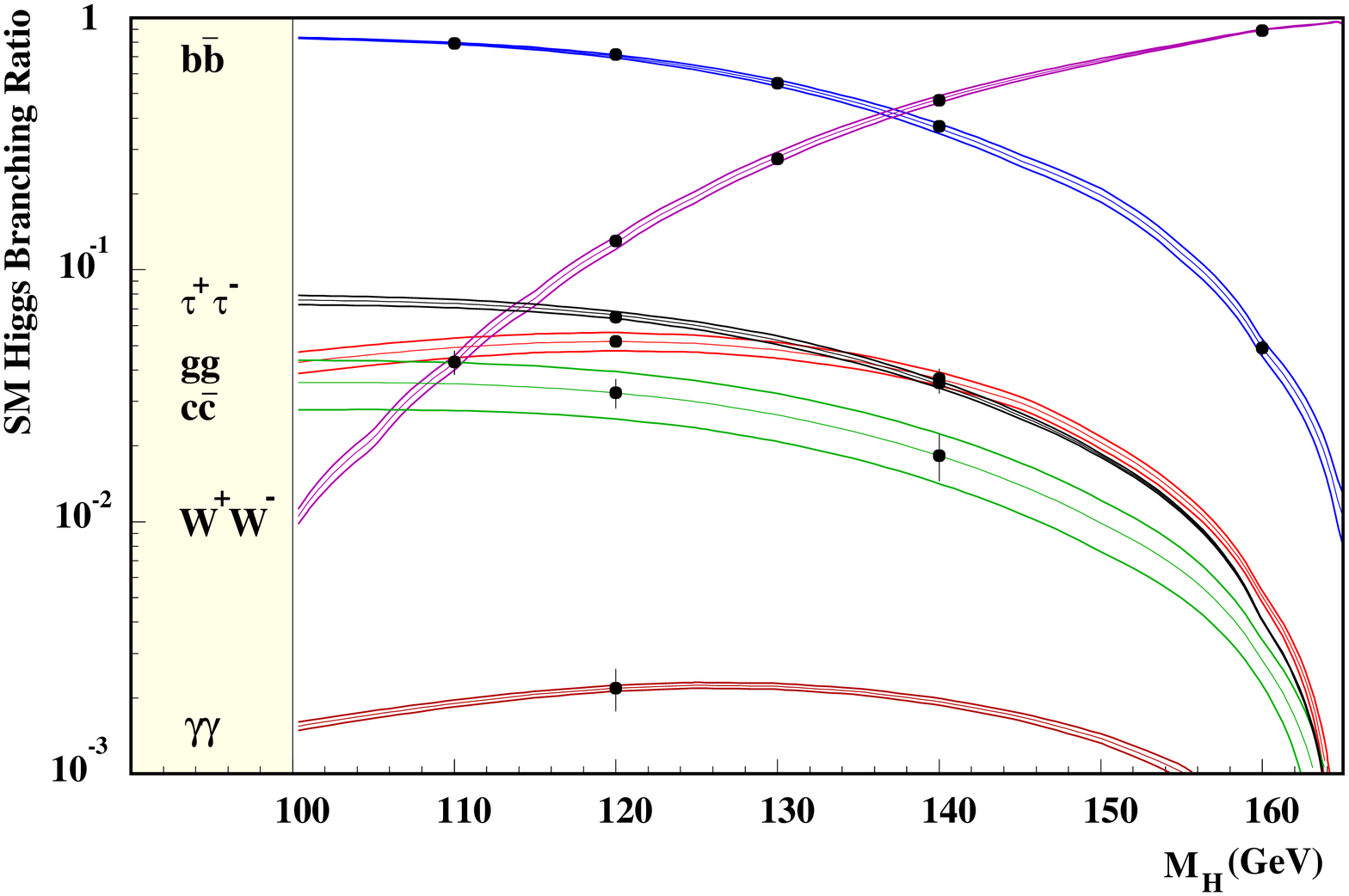}
}
\epsfclipoff
\vspace*{8pt}
\caption{\label{fig:two} The predicted SM Higgs boson branching
ratios. Points with error bars show the expected experimental accuracy,
while the lines show the estimated uncertainties on the SM
predictions. (from Ref.~[1]).} 
\end{figure}
Of particular interest at the ILC is the precise measurement of the Higgs
boson couplings to fermions and the weak bosons (once a Higgs 
candidate particle has been found). At the LHC, these couplings can be
measured with a precision of ${\cal O}(10\%)$ at best\cite{Duhrssen:2004cv}. 
At the ILC it will be possible to determine the Higgs boson couplings
with an accuracy of a few
percent\cite{Aguilar-Saavedra:2001rg}. Figure~\ref{fig:two} shows the
predicted branching ratios for various Higgs decays, the 
expected experimental precision, and the current theoretical
uncertainties as a function of the Higgs boson mass.
If the Higgs boson mass is $\leq 140$~GeV, 
it may also be possible to measure the Higgs boson self-coupling,
$\lambda_{HHH}$, at the
ILC, and thus to directly probe the Higgs
potential\cite{Castanier:2001sf}. At the LHC, $\lambda_{HHH}$ can 
be probed in this mass range\cite{Baur:2003gp} only once the luminosity has
been been upgraded to ${\cal L}=10^{35}\,{\rm cm^{-2}\, s^{-1}}$. In
order to probe the Higgs boson couplings at the ILC with the advertised
accuracy, the one-loop electroweak radiative corrections to $e^+e^-\to ZH$, 
$e^+e^-\to \nu\bar\nu H$, $e^+e^-\to e^+e^- H$, $e^+e^-\to t\bar tH$,
$e^+e^-\to ZHH$,  
$e^+e^-\to \nu\bar\nu HH$, and $e^+e^-\to e^+e^- HH$ are needed. Thanks
to new automated tools which I will discuss in more detail in the
following Section, the one-loop electroweak radiative corrections to all
these processes except $e^+e^-\to e^+e^- HH$ have been calculated in the
last few years\cite{ilc_calc}. 

\begin{figure}[t]
\begin{center}
\begin{tabular}{cc}
\epsfxsize=5.8cm
\epsffile{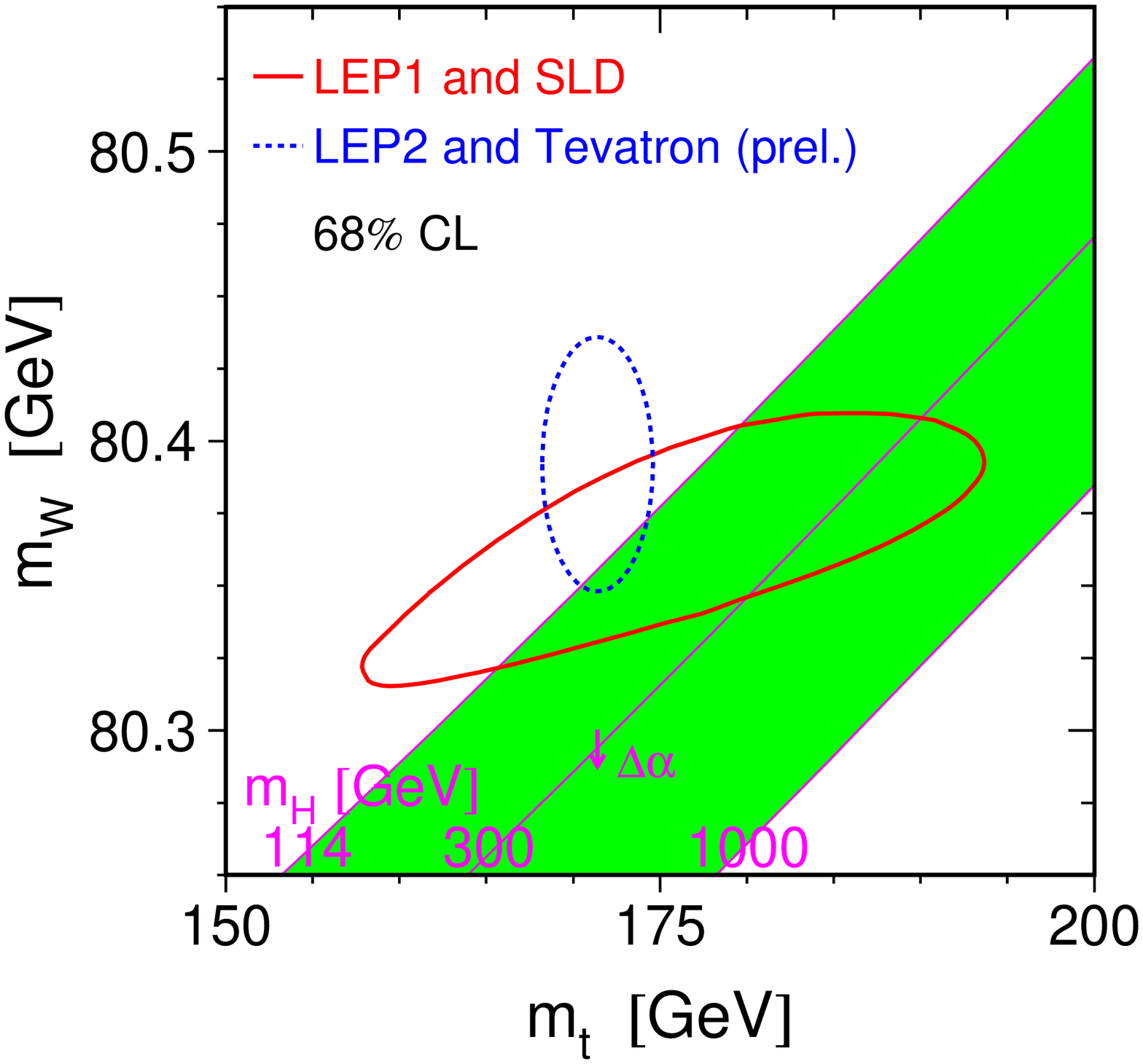} &
\epsfxsize=5.8cm
\epsffile{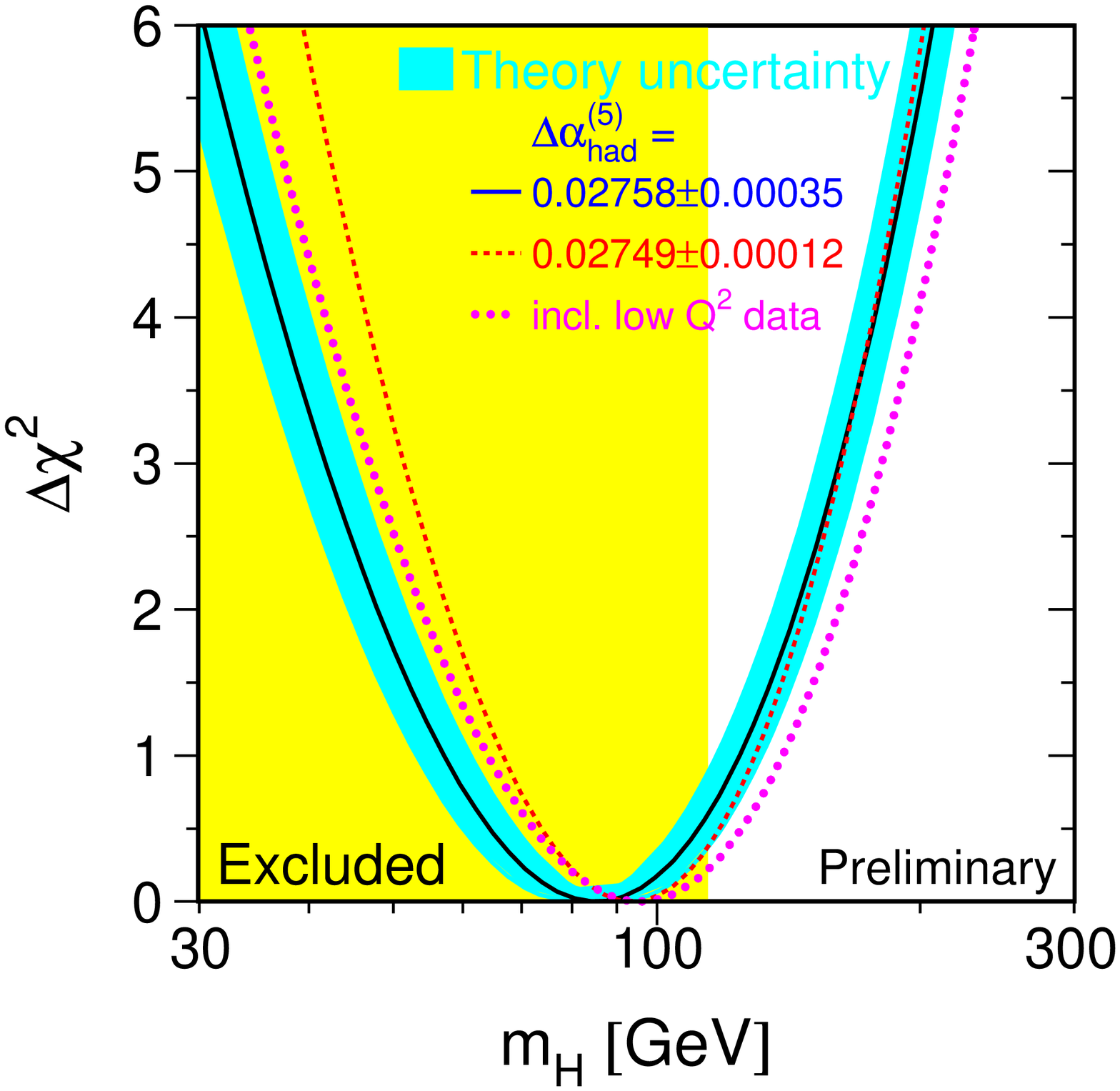}
\end{tabular}
\end{center}
\vspace*{-8pt}
\caption{\label{fig:three} Left pane: the experimentally allowed regions in
the $m_W-m_t$ plane are shown together with the SM predictions.
Right pane: $\Delta\chi^2-\chi^2-\chi^2_{min}$ vs. $m_H$ curve. The line
is result of a fit to all electroweak data, and the blue band represents
an estimate of the theoretical error due to missing higher order
corrections. The vertical yellow band represents the 95\% CL bound from
searches at LEP2\protect\cite{Barate:2003sz}.}  
\end{figure}

Other electroweak observables which can be precisely measured at the ILC
are the $W$ mass, $m_W$, and the effective weak mixing angle,
$\sin^2\theta_{eff}$. The one-loop electroweak corrections to these
parameters depend quadratically on the top quark mass, $m_t$, and
logarithmically on the Higgs boson mass, $m_H$. Thus, measuring $m_W$ or
$\sin^2\theta_{eff}$ and $m_t$ makes it possible to extract information
on $m_H$. The regions currently allowed by LEP1 and SLC data, and LEP2
and Tevatron data in the $m_W-m_t$ plane, together
with the SM prediction are shown in Fig.~\ref{fig:three}a. 
Figure~\ref{fig:three}b shows the $\Delta\chi^2$ curve as a function of
$m_H$. The blue band represents
an estimate of the theoretical error due to missing higher order
corrections. It is dominated by the theoretical uncertainty on the effective
weak mixing angle for which the full two-loop electroweak corrections are now
known\cite{Awramik:2006uz}. Still, the remaining theoretical uncertainty
is
\begin{equation}
\delta\sin^2\theta_{eff}^{theor}\approx (4-5)\times 10^{-5},
\end{equation}
which is only a factor $3-4$ smaller than the experimental uncertainty
of $\delta\sin^2\theta_{eff}^{exp}=0.00017$\cite{Group:2006mx}. I will
come back to the measurement of 
$\sin^2\theta_{eff}$ and the $W$ mass at the ILC in Sec.~3.2.

The reminder of this review is organized as follows. In Sec.~2, I
briefly outline the framework of higher order calculations in
perturbation theory and briefly discuss the tools which currently
available for performing these calculations. In Sec.~3, I discuss recent
results which are relevant for the LHC and ILC. Section~4 contains a
summary and outlook.

\section{Tools for Loop Calculations}

\begin{figure}[t]
\begin{center}
\begin{tabular}{ccc}
\epsfxsize=3.8cm
\epsfclipon
\epsffile[150 550 450 740]{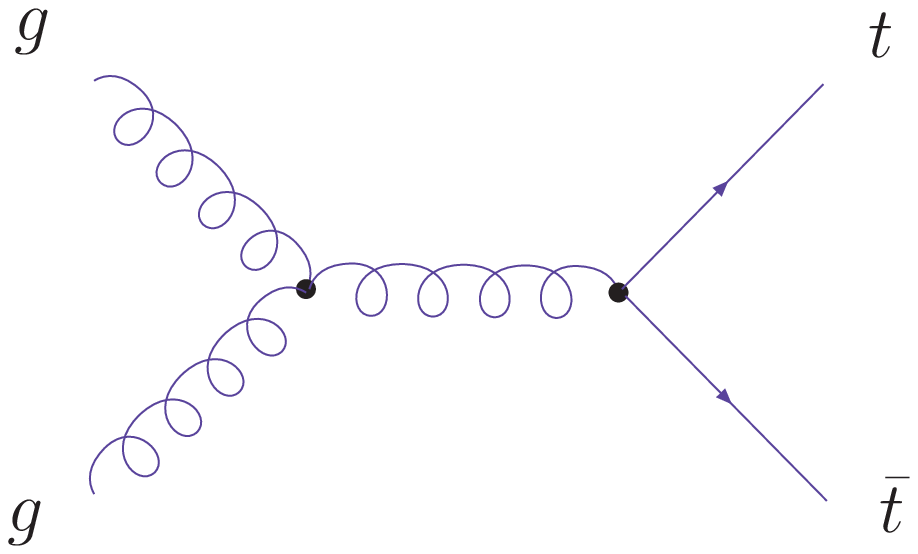} &
\epsfxsize=2cm
\epsffile{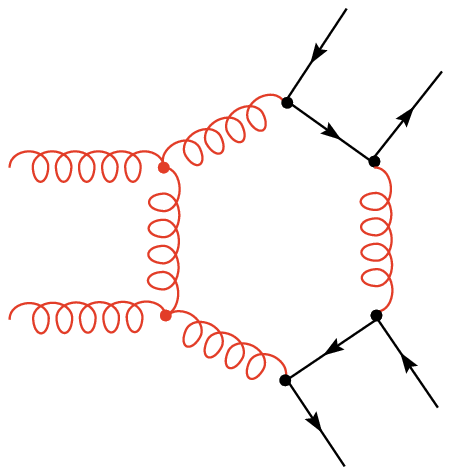} &
\epsfxsize=3.8cm
\epsffile{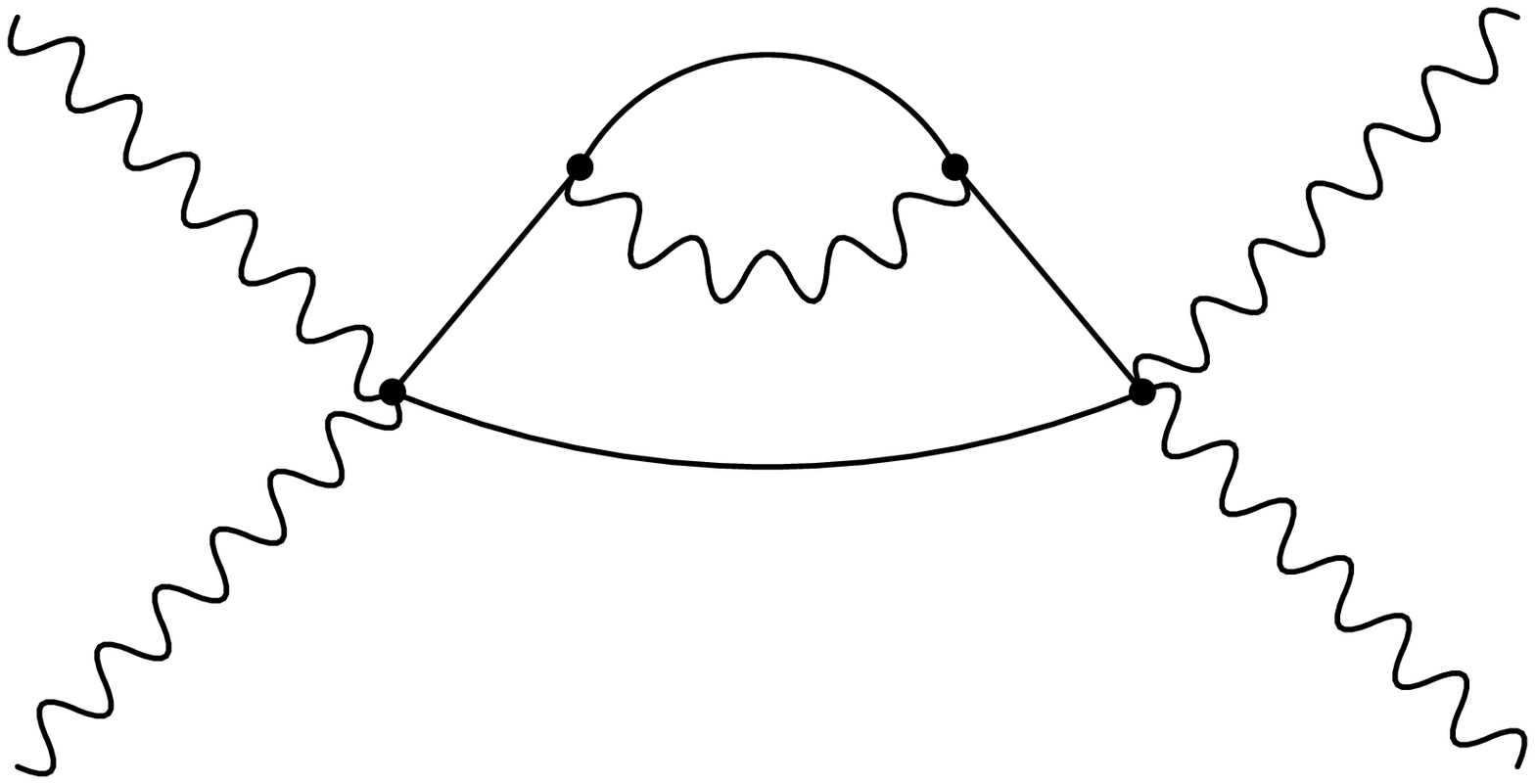}
\end{tabular}
\end{center}
\vspace*{-8pt}
\caption{\label{fig:four} Examples for tree level (left), one-loop
(middle) and two-loop (right) Feynman diagrams.}  
\end{figure}

The theoretical framework for loop calculations in high energy physics
is perturbation theory. In perturbation theory, the observable of
interest is expanded in powers of a (small) coupling constant. The
lowest order (LO) terms correspond to tree level Feynman diagrams, the
NLO corrections involve one-loop diagrams, and so on. Examples for tree
level, one-loop and two-loop diagrams are shown in Fig.~\ref{fig:four}. 

The general strategy of a loop calculation is best illustrated using a
simple process such as 2~jet production in $e^+e^-$ collisions,
$e^+e^-\to q\bar q$, as an example. 
In order to compute the NLO QCD
corrections to $e^+e^-\to q\bar q$, one needs to calculate the one-loop
corrections to this process, and the tree level process $e^+e^-\to q\bar
qg$. Both occur at the same order in perturbation theory. Due
to soft and collinear divergencies, the cross section for $e^+e^-\to
q\bar q$ at one-loop and for $e^+e^-\to q\bar qg$ each
diverges; however, their sum is finite and represents the physical cross
section for 2~jet production in $e^+e^-$ collisions at NLO in QCD. 

Computing the $e^+e^-\to q\bar qg$ cross section is the easy part of
the calculation. Tree level calculations are technically
straightforward, and a number of automatic programs exists which greatly
simplify the task. The most general and flexible tools are {\tt
MadEvent}\cite{Maltoni:2002qb}, {\tt Grace}\cite{Yuasa:1999rg}, {\tt
CalcHEP}\cite{Pukhov:2004ca}, {\tt
CompHEP}\cite{Boos:2004kh}, and {\tt WHIZARD}\cite{kilian} which 
allow the user to completely specify the process he/she wishes to
calculate. The flexibility of these programs comes at the price of
speed. {\tt ALPGEN}\cite{Mangano:2002ea}, on the other hand, is
extremely fast, but works only
for a selected set of processes which are hard coded into the
program. {\tt MadEvent}, however, can be run in parallel on several
machines, and a future version of {\tt CompHEP} may be able to do the
same\cite{Kryukov:2004md}. This (partially) compensates the relative
slowness of these programs. 

In addition to calculating the cross sections
for SM processes, most programs are now able to also compute
beyond-the-SM (eg. supersymmetric) processes.
The capabilities of programs which automatically calculate tree level
processes is only limited by the computing
power available to the user. For multi-particle final states, thousands
of Feynman diagrams may contribute; the number of diagrams grows
factorially with the number of final state particles. For the process
$e^+e^-\to W^+W^-\bar bbjj$, for example, there are 4896 Feynman
diagrams. The numerical evaluation of matrix elements takes
progressively more time the more Feynman diagrams contribute.

The calculation of the one-loop corrections to $e^+e^-\to q\bar q$ is
considerably more involved. The general strategy of a loop calculation
is outlined in Table~\ref{tab:one}.
\begin{table}[t]
\caption{\label{tab:one}Outline of a loop calculation.}
\begin{tabular}{|l|l|}
\hline
draw all possible diagrams & topological task \\
which particles run in given diagram & combinatorial task\\
translate diagrams into formulas via Feynman rules & database look-up\\
contract Lorentz indices; take traces & algebraic manipulation\\
reduce to known/master integrals & algebraic manipulation\\
cancel IR and/or UV singularities & algebraic manipulation\\
translate output into computer program & programming\\
run program & wait, drink coffee\\
\hline
\end{tabular}
\end{table}
Even for moderately complicated processes such as $e^+e^-\to 4$~fermions
the number of Feynman diagrams which has to be calculated can be
extremely large (${\cal O}(10^4)$). Additional complications arise from
large cancellations which occur between certain Feynman diagrams. This
requires extra care with the numerical implementation. 

\begin{figure}[t]
\epsfxsize=12.cm
\epsfclipon
\centerline{
\epsffile[0 0 792 612]{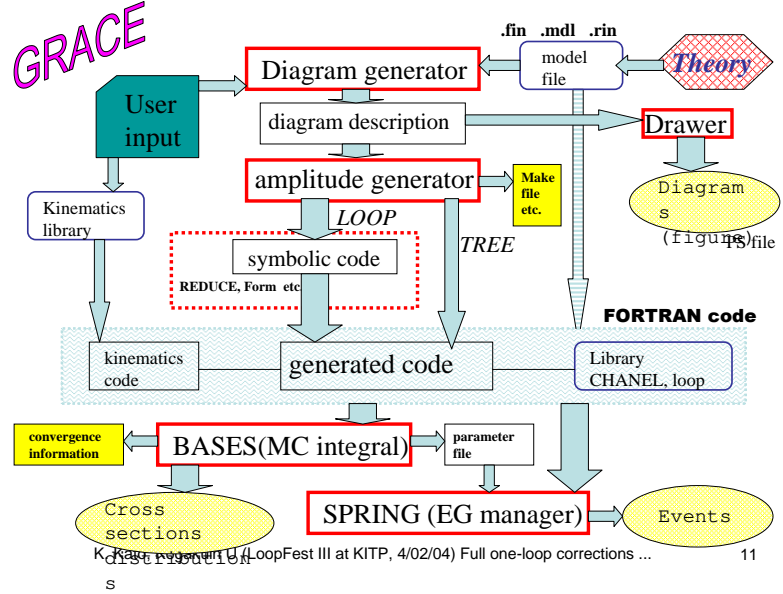}
}
\epsfclipoff
\vspace*{-8pt}
\caption{\label{fig:five} Flow diagram of the {\tt Grace/1-loop}
package. (from Ref.~[23]).} 
\end{figure}

Because of the complexity of loop calculations, automatic tools are
essential to accomplish the goal. So far, program packages for automated
loop calculations only exist for electroweak one-loop corrections. The
{\tt Grace/1-loop}\cite{Belanger:2003sd} package has been used
successfully in calculating a number of processes relevant for the
ILC. Another popular set of semi-automatic tools are {\tt
Feynarts\cite{Hahn:2000kx}, FeynCalc\cite{Mertig:1990an},
FormCalc\cite{Hahn:1998yk}} and {\tt
LoopTools}\cite{Hahn:1998yk}. Finally, {\tt
Diana}\cite{Tentyukov:2003bx} is an automatic tool for generating the
Feynman diagrams which contribute to a given process and a given order
in perturbation theory. {\tt Diana} also produces the input needed for
programs which evaluate the traces of Dirac-matrices, such as {\tt
FORM}\cite{Vermaseren:2000nd}. Not surprisingly, the structure of these
programs is fairly complex. As an example, I show the flow diagram of
the {\tt Grace/1-loop} package in Fig.~\ref{fig:five}.
Packages for automated calculations of one-loop QCD corrections are
currently under development ({\tt Grace-QCD}\cite{Kurihara:2006kt} and
{\tt Samper}\cite{walter}). 

\begin{figure}[t]
\epsfxsize=12.cm
\epsfclipon
\centerline{
\epsffile[-100 150 595 662]{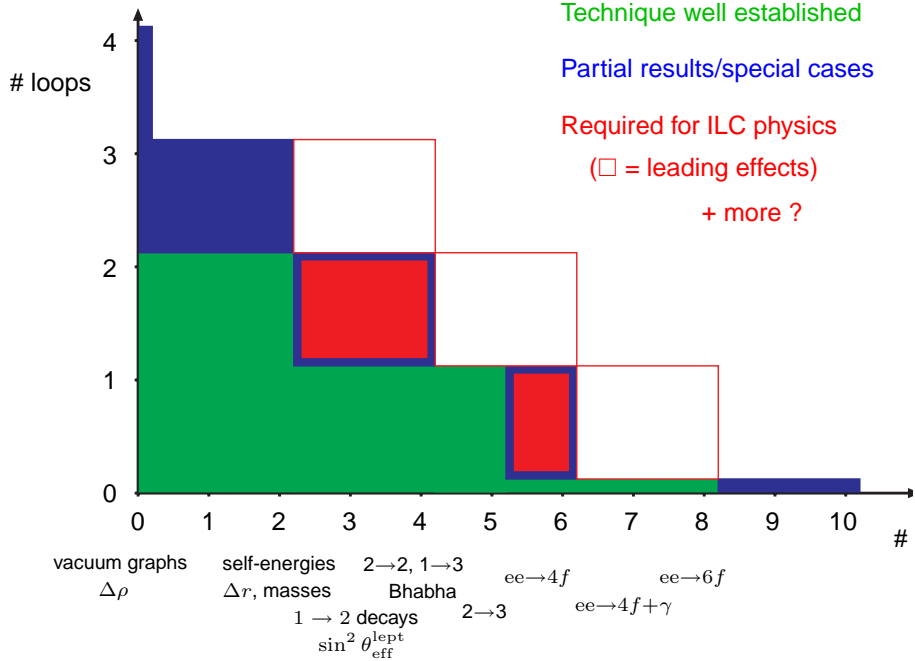}
}
\epsfclipoff
\vspace*{-8pt}
\caption{\label{fig:six} The present state of art of calculations of
higher order corrections for ILC processes (from Ref.~[26]).} 
\end{figure}

The complexity of loop calculations rapidly increases with the number of
loops, and with the number of particles in the final state. On the other
hand, the cross
section of processes falls rather quickly with the number of final state
particles. Thus, the requirements on the theoretical accuracy for $2\to
n$, $n>2$, processes are less than for $2\to (n-1)$ processes, ie. the
order in perturbation theory up to which one needs to calculate
cross sections decreases with increasing number of particles in the
final state. Figure~\ref{fig:six} shows the loop order
(which is equivalent to the order of perturbation theory) which one
needs to calculate for processes of interest at the ILC as a function of
the number of particles in the final 
state. The figure also shows the current status of calculations of
higher order corrections for these processes.

\section{Recent Results}

In this Section I discuss some recent results in precision calculations
relevant for LHC and ILC processes.

\subsection{One-loop Corrections for LHC Processes}

Multijet production is an important background for many processes of
interest at the LHC. The NLO QCD corrections to 2~jet and 3~jet
production have been known for several years\cite{twojet,threejet}. In
contrast, the calculation of the NLO QCD corrections to 4~jet production
is just beginning. The most complicated contribution to the one-loop
corrections to $pp\to 4$~jet originates from $gg\to gggg$. The one-loop
corrections to this sub-process were recently calculated in
Ref.~[29]. Approximately 10,000 Feynman diagrams contribute to $gg\to
gggg$ at the one-loop level. To obtain numerical results, a new
promising method was used which semi-numerically evaluates loop
integrals\cite{Ellis:2005zh}. 

\begin{figure}[t]
\epsfxsize=12.cm
\epsfclipon
\centerline{
\epsffile{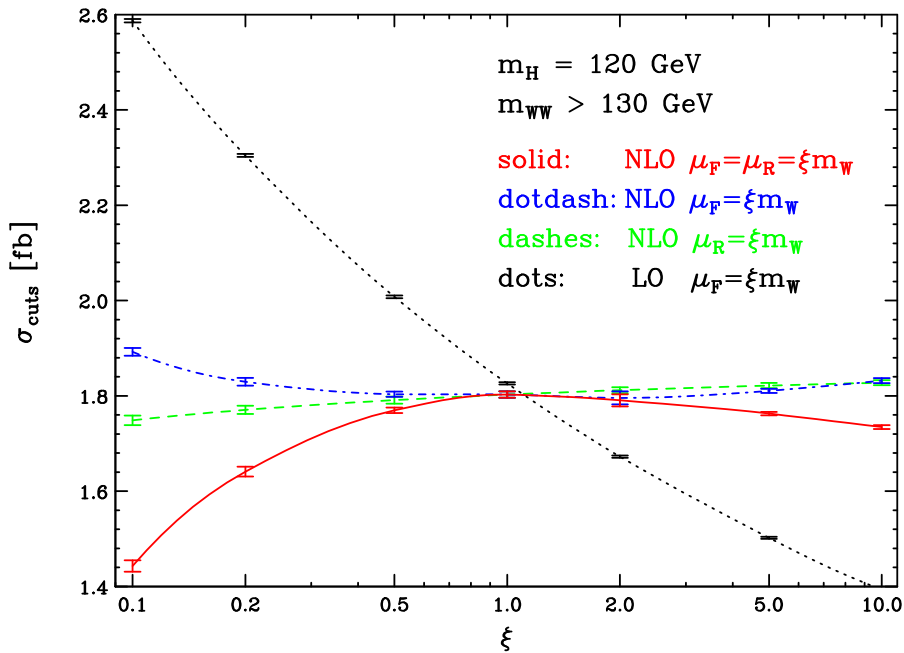}
}
\epsfclipoff
\vspace*{-8pt}
\caption{\label{fig:seven} The scale dependence of the $qq'\to
W^+W^-qq'$ cross section for $pp$ collisions with $\sqrt{s}=14$~TeV at
LO and NLO (from Ref.~[31]).}  
\end{figure}

In Sec.~1, I noted that calculating higher order QCD corrections usually
reduces the sensitivity of the cross section on the renormalization and
factorization scales, $\mu_R$ and $\mu_F$. A recent
calculation\cite{Jager:2006zc} of the 
NLO QCD to $W^+W^-$ production via vector boson fusion (VBF), $qq'\to
W^+W^-qq'$, nicely illustrates this point. $W$ pair production via VBF
is one of the most important Higgs discovery channels at the
LHC\cite{Kauer:2000hi}. While the LO $qq'\to W^+W^-qq'$ cross section
varies very strongly with $\mu_R$ and $\mu_F$, the NLO cross section is
almost independent of the 
choice of scale over a wide range (see Fig.~\ref{fig:seven}).

As mentioned in Sec.~2, one has to include real quark/gluon radiation
diagrams when calculating higher order corrections in QCD in order to
obtain a finite cross section. Individually, the cross sections obtained
from virtual and real corrections are infinite due to soft and collinear
divergencies. This also happens when calculating QED radiative
corrections. The soft and collinear divergencies are due to the
vanishing mass of the QCD and QED gauge bosons. 

In contrast to QCD and QED, the electroweak gauge bosons are massive and
act as infrared regulators. Cross sections for real and virtual
weak corrections thus are separately finite. The virtual weak
corrections turn out to become large and negative at high energies, due
to the presence of Sudakov-like logarithms of the form $(\alpha/\pi)\log^2(\hat
s/m_{W,Z}^2)$, where $\hat s$ is the squared parton center of mass 
energy, and $m_{W,Z}$ is the mass of the $W$ or $Z$ boson. For
$\sqrt{\hat s}\geq 1$~TeV, the ${\cal O}(\alpha)$ one-loop EW radiative
corrections can easily become larger in magnitude than the ${\cal
O}(\alpha_s)$ QCD corrections. 

\begin{figure}[t]
\epsfxsize=12.cm
\epsfclipon
\centerline{
\epsffile[50 330 300 560]{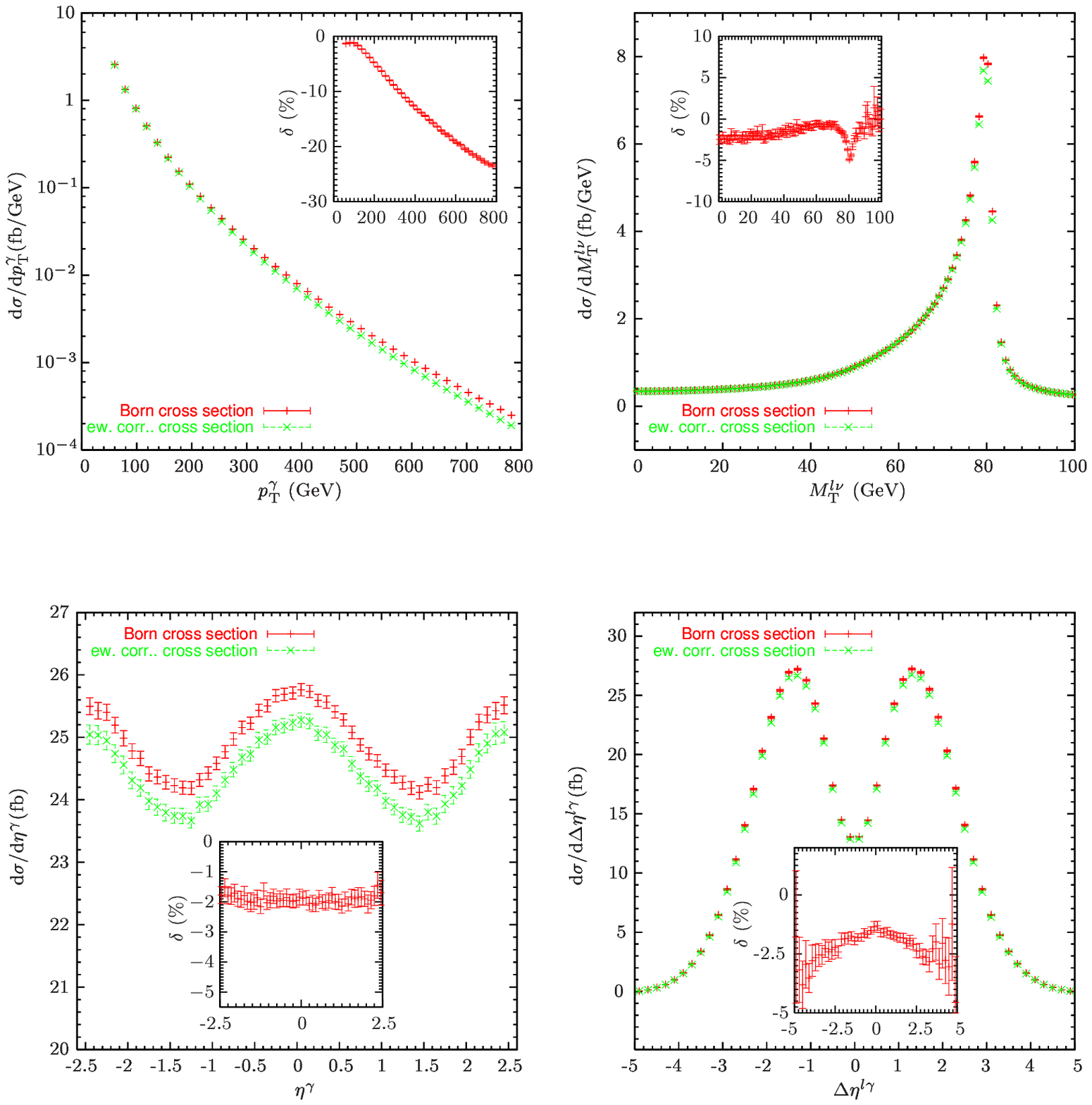}
}
\epsfclipoff
\vspace*{-8pt}
\caption{\label{fig:eight} The photon transverse momentum distribution
in $pp\to W\gamma\to e\nu\gamma$ for $\sqrt{s}=14$~TeV at LO and
including weak virtual 
corrections (from Ref.~[50]). The inset shows the relative corrections.} 
\end{figure}

The Sudakov-like logarithms originate from collinear and infrared
divergences which would be present in the limit of vanishing $W$ and $Z$
masses and are well
understood\cite{Ciafaloni:1998xg,Ciafaloni:2000df,Ciafaloni:2000gm,Denner:2000jv,Kuhn:1999de,Melles:2001ye}.
The appearance of large logarithms in one-loop weak corrections has
recently been demonstrated in a number of explicit calculations. For
hadron colliders, 
the ${\cal O}(\alpha)$ virtual weak corrections to inclusive
jet\cite{Moretti:2005ut}, isolated photon\cite{Kuhn:2005gv,Maina:2004rb},
$Z+1$~jet\cite{Maina:2004rb,Kuhn:2004em},
Drell-Yan\cite{Baur:2001ze,kramer,Baur:2004ig,Arbuzov:2005dd,Zykunov:2005tc,CarloniCalame:2006zq},
di-boson\cite{Accomando:2001fn,wgam,Accomando:2004de}, $\bar
tt$\cite{Kuhn:2005it,mnr,Beenakker:1993yr,Bernreuther:2006vg}, and single top
production~\cite{Beccaria:2006ir,Beccaria:2006dt,Ciafaloni:2006qu} have
been calculated. As an example, I show the virtual
weak corrections for the photon transverse momentum distribution in
$pp\to W\gamma\to e\nu\gamma$ in Fig.~\ref{fig:eight}. They strongly
increase in magnitude with increasing photon transverse momentum and
reach about $-25\%$ at $p_T(\gamma)=800$~GeV.
In almost all of the calculations where large logarithms appear in
one-loop weak corrections, weak boson emission
diagrams have not been taken into account, although
they contribute at the same order in perturbation theory as the one-loop
corrections and often contribute substantially to the NLO electroweak
cross section\cite{Baur:2006sn}. 

\subsection{Recent Calculations relevant for the ILC}

As mentioned in Sec.~1, it will be possible to measure the effective
weak mixing angle and the $W$ mass at the ILC. The ILC detectors can be
calibrated by measuring the $Z$ boson mass and comparing the result to the
value obtained at LEP1. This requires operation of the ILC at the $Z$
peak, which also offers a chance to determine $\sin^2\theta_{eff}$. 
The $W$ mass can either be measured by directly reconstructing the $W$
bosons in $e^+e^-\to W^+W^-$, or by measuring the the $W$ pair cross
sections in the threshold region ($\sqrt{s}\approx 161$~GeV). The latter
method promises to be more precise at the ILC.

As noted before, the full two loop electroweak corrections to
the effective weak mixing angle were recently
calculated\cite{Awramik:2006uz}. This, however, will not be sufficient
for the ILC. At the ILC one hopes to measure $\sin^2\theta_{eff}$ with a
precision of~\cite{Hawkings:1999ac}
\begin{equation}
\delta\sin^2\theta_{eff}^{exp}=1.3\times 10^{-5},
\end{equation}
which about a factor three smaller than the current theoretical
uncertainty from unknown higher order corrections (see Eq.~(1)). For a
measurement of the effective weak mixing at the ILC, one thus has to
calculate the 3-loop ${\cal O}(\alpha_s\alpha^2)$ corrections. 

The $W$ mass can be measured with a precision of about 7~MeV in a
threshold scan at the ILC\cite{Monig:2003ze}. This means that the
$e^+e^-\to 4$~fermion cross section has to be known with a precision
of~\cite{Stirling:1995xp} 
\begin{equation}
{\Delta\sigma\over\sigma}\approx 5\times 10^{-4}
\end{equation}
in the threshold region. The uncertainty of the LO $e^+e^-\to 4$~fermion
cross section at $\sqrt{s}=161$~GeV is
approximately\cite{Grunewald:2000ju} 
\begin{equation}
\left({\Delta\sigma\over\sigma}\right)_{LO}=0.014.
\end{equation}
For a $W$ mass measurement from a threshold scan at the ILC, one
therefore needs the full ${\cal O}(\alpha)$ electroweak radiative
corrections to $e^+e^-\to 4$~fermions. These corrections have recently
been calculated\cite{Denner:2005es}. A major complication which had to
be overcome in this calculation is how to include finite $W$ width
effects while maintaining gauge invariance. Including the $W$ width in
the $W$ propagator corresponds to a resummation of
the imaginary part of the $W$ vacuum polarization, and is
essential in the threshold region for obtaining a realistic prediction
of the cross section. Since only a subset
of the Feynman diagrams which contribute to $e^+e^-\to 4$~fermions is
resummed in this procedure, this will break 
gauge invariance. The gauge invariance problem was solved in Ref.~[64]
by using the complex mass scheme and complex renormalization. 

\begin{figure}[t]
\begin{center}
\begin{tabular}{cc}
\epsfclipon
\epsfxsize=6.2cm
\epsffile[140 430 400 710]{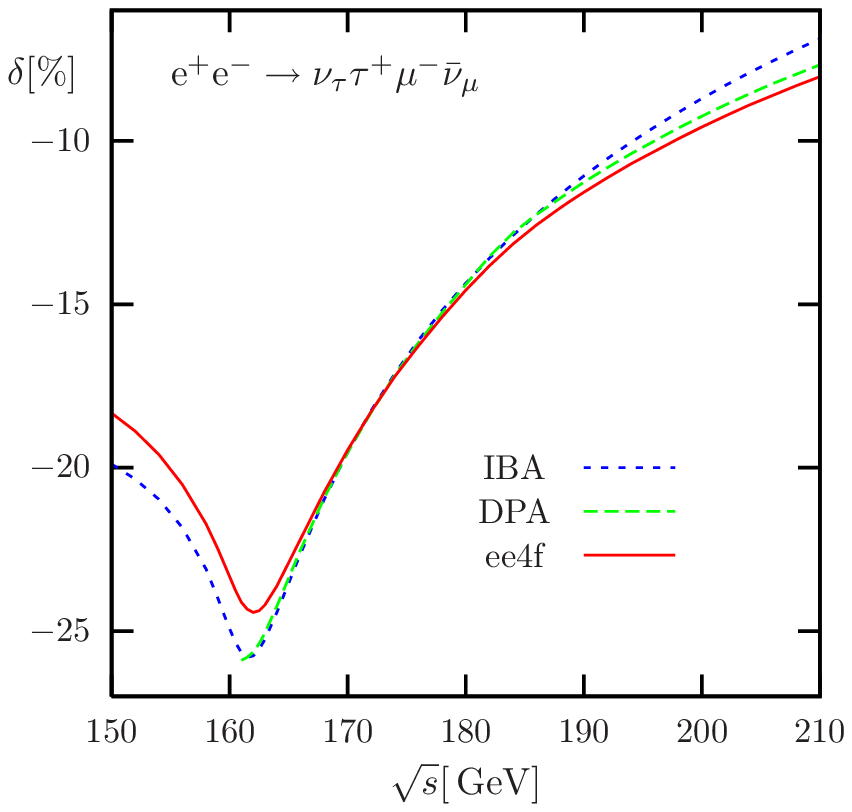} &
\epsfxsize=6.2cm
\epsffile[170 430 430 530]{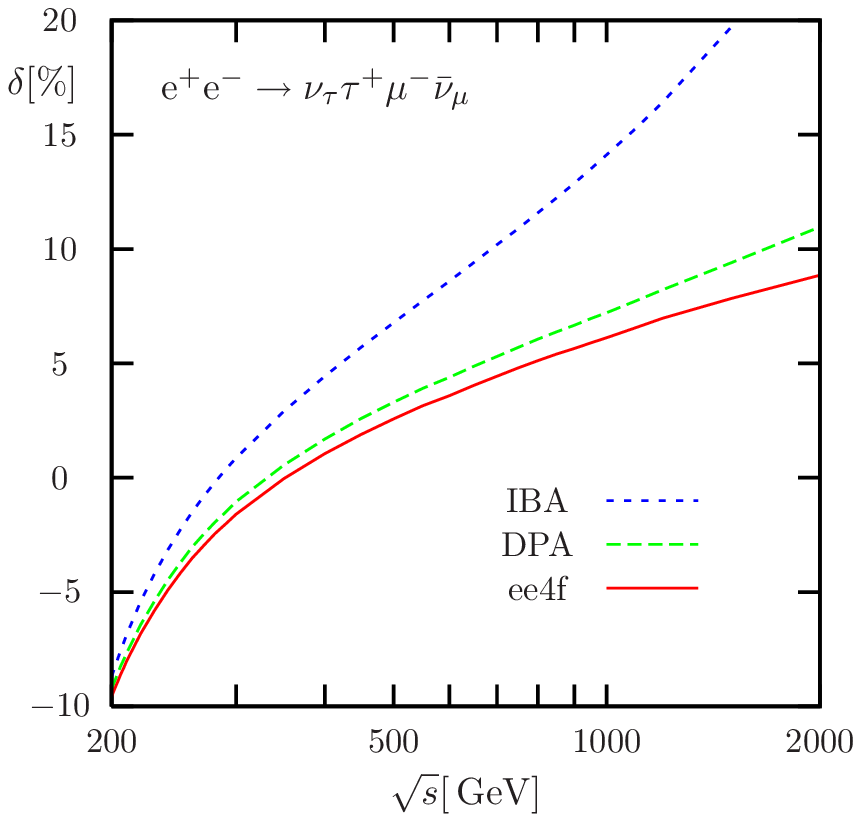}
\epsfclipoff
\end{tabular}
\end{center}
\vspace*{-28pt}
\caption{\label{fig:nine} The relative ${\cal O}(\alpha)$ electroweak
radiative corrections to $e^+e^-\to\tau^+\nu_\tau\mu^-\bar\nu_\mu$ as a
function of the center of mass energy, $\sqrt{s}$. Shown are the results
for the improved Born approximation (IBA), the ${\cal O}(\alpha)$
electroweak radiative corrections in the double pole approximation
(DPA), and the full ${\cal O}(\alpha)$ electroweak radiative corrections
(ee4f) (from Ref.~[64]).}
\end{figure}

A main technical challenge in the calculation of the full ${\cal O}(\alpha)$
electroweak radiative corrections to $e^+e^-\to 4$~fermions is the
reduction of hexagon diagrams to box diagrams which, employing
conventional methods, leads to numerical instabilities. Ref.~[64]
overcame this problem by using Cayley
determinants\cite{Denner:2005nn}. To illustrate the results obtained in
Ref.~[64], I show the relative corrections to the
$e^+e^-\to\tau^+\nu_\tau\mu^-\bar\nu_\mu$ cross section in
Fig.~\ref{fig:nine}. 
In the $W$ threshold region, the difference between the full and
approximate ${\cal O}(\alpha)$ electroweak radiative corrections is seen
to be approximately 2\%. 

The calculation of the full ${\cal O}(\alpha)$ electroweak radiative
corrections to $e^+e^-\to 4$~fermions represents a major step
forward. Many other 
calculations of radiative corrections to $2\to 4$ processes are now
feasible. 
However, the ${\cal O}(\alpha)$ corrections to $e^+e^-\to 4$~fermions
may not be sufficient for a $W$ mass measurement with a precision of
7~MeV at the ILC. Next-to-leading logarithmic electromagnetic corrections
of order 
$(\alpha/\pi)^2\log(m_e^2/s)$, where $m_e$ is the electron mass, and
higher order effects associated with the Coulomb singularity may
modify the $e^+e^-\to 4$~fermion cross section by ${\cal
O}(10^{-3})$\cite{Denner:2005es,Fadin:1995fp,Bardin:1993mc}. These
corrections still have to be calculated.

\subsection{Recent Results from the Two-loop Frontier}

Enormous progress in calculating two-loop corrections has been made in
the last few years. Key developments have been the algebraic reduction
of loop integrals 
to master integrals by integration by parts and Lorentz invariance
identities, and the calculation of master integrals using the 
Mellin-Barnes technique, differential equations and numerical
techniques\cite{Anastasiou:2000mf}. This has lead to a number of results
for explicit two-loop amplitudes such as 2~jet production at the
LHC\cite{glover}, $e^+e^-\to 3$~jets\cite{Garland:2002ak}, and
$e^+e^-\to e^+e^-$\cite{Bern:2000ie}. Of course, for a full calculation
of the NNLO corrections to these processes, the two-loop amplitudes
have to be combined with the relevant one-loop $2\to 3$ and $2\to 4$
amplitudes, and the tree level $2\to4$ and $2\to 5$ amplitudes. This
requires the development of a suitable subtraction method for the soft
and collinear divergencies which appear in the calculation. Several 
promising
techniques\cite{Gehrmann-DeRidder:2005cm,Anastasiou:2003gr,Binoth:2004jv}
are currently pursued.

For some processes, such as $H\to\gamma\gamma$\cite{Anastasiou:2005qj},
$W\to\ell\nu$ and $Z\to\ell^+\ell^-$\cite{Anastasiou:2003ds} production
in hadronic collisions, the fully differential NNLO QCD cross section is
already available. This makes it 
possible to study in detail how QCD corrections affect the experimental
acceptances for these processes. The $W$ (and $Z$) boson cross
section can be used as a luminosity monitor at the
LHC\cite{Dittmar:1997md}. This requires 
the theoretical uncertainty on the cross section to be below
1\%. Knowledge of the NNLO QCD corrections to the fully differential $W$
cross section is an essential ingredient to achieve this
goal. $H\to\gamma\gamma$ is an important Higgs discovery channel if
$m_H<140$~GeV\cite{Buscher:2005re}. 

\subsection{New Theoretical Developments}

Recent progress in the analytical computation of tree-level\cite{tree}
and massless one-loop\cite{oneloop} gauge theory amplitudes  provides a
promising alternative to the techniques used so far. This work,
including new methods 
based on twistor-space string theories\cite{Witten:2003nn}, has led to
compact expressions and recursion relations that promise a faster
numerical evaluation of differential cross sections. The next steps 
in bringing this approach to fruition 
are to generalize the results for
massless one-loop diagrams to the massive case, and to build
parton-level MC programs for processes of interest.  

\subsection{Recent Results for Supersymmetric Theories}

In addition to precision calculations in the framework of the SM, many
such calculations have been performed for supersymmetric theories. It is
impossible to mention all results here so I concentrate on a few
selected calculations. 

The one-loop radiative corrections to the 
$W$ mass in the minimal supersymmetric SM (MSSM) have been known for more
than 10~years\cite{dr1lA,dr1lB,MWMSSM1LA,MWMSSM1LB,pierce}. More
recently, the Yukawa corrections to the
$\rho$-parameter\cite{Heinemeyer:2002jq} and the Higgs masses and widths
in MSSM\cite{Frank:2006yh} have been computed.
For a recent review of electroweak precision observables in the MSSM see
Ref.~[88]. 
The one-loop corrections to chargino and neutralino pair production in
$e^+e^-$ collisions have been evaluated in Refs.~[89] and~[90]. New
tools for supersymmetric processes include {\tt Prospino 2.0} which
computes next-to-leading order cross sections for the production of
supersymmetric particles at hadron colliders\cite{Beenakker:1996ch},
{\tt Sdecay} which calculates the decay widths and branching ratios of
all supersymmetric particles in the MSSM\cite{Muhlleitner:2003vg}, and
{\tt SUSY-Madgraph} which generates complete tree level matrix elements
for the production of supersymmetric particles, including decays and
spin correlations\cite{Cho:2006sx}.

\section{Summary and Outlook}

Accurate theoretical predictions for SM and beyond-the-SM processes are
needed in order to correctly interpret data from the LHC and ILC. In the
last few years enormous progress has been made in developing new
techniques for loop calculations and new tools for tree level
calculations for complex final states. The one-loop corrections for
essentially all $2\to 2$ processes of interest are known. Thanks to
automated tools, calculations of one-loop corrections to $2\to 3$
processes have become fairly routine. The frontier in one-loop
corrections now are $2\to 4$ processes, such as $e^+e^-\to 4$~fermions.

Although much has been
accomplished, there remain significant challenges. For example, the one
loop corrections 
for $2\to 5$ processes such as $pp\to WWWjj$, which is a background
relevant for the determination of the Higgs boson self-coupling in Higgs
pair production at the LHC\cite{Baur:2002rb}, have not been tackled
yet. While the two loop corrections 
for many $2\to 2$ processes are known, these have yet to be combined with
the one-loop amplitudes of the relevant $2\to 3$, and the tree level
amplitudes of the corresponding $2\to 4$ processes in order to obtain
physical cross sections. The fully differential cross section including
NNLO corrections is only known for a few processes.
Beyond the two-loop level, calculations and tools are still in their
infancy. 

Approximately 85\% of the work done on precision calculations is carried out
in Europe or Asia, as evidenced by the references included in this
review. Clearly, a stronger role of the Americas in this field which is
of vital importance for the LHC and ILC, and thus more regional balance,
is desirable. I hope that with the LHC approaching this will happen.

\section{Acknowledgments}
This research was supported by the 
National Science Foundation under grant No.~PHY-0456681.


\begin{thebibliography}{99}
%
\bibitem{Aguilar-Saavedra:2001rg}
  J.~A.~Aguilar-Saavedra {\it et al.}  [ECFA/DESY LC Physics Working Group],
  arXiv:hep-ph/0106315.
%
\bibitem{Anastasiou:2003ds}
  C.~Anastasiou, L.~J.~Dixon, K.~Melnikov and F.~Petriello,
  Phys.\ Rev.\ D {\bf 69}, 094008 (2004)
  [arXiv:hep-ph/0312266];
K.~Melnikov and F.~Petriello,
  Phys.\ Rev.\ Lett.\  {\bf 96}, 231803 (2006)
  [arXiv:hep-ph/0603182] and
  Phys.\ Rev.\ D {\bf 74}, 114017 (2006)
  [arXiv:hep-ph/0609070].
%
\bibitem{Duhrssen:2004cv}
  M.~D\"uhrssen, S.~Heinemeyer, H.~Logan, D.~Rainwater, G.~Weiglein and
D.~Zeppenfeld, 
  Phys.\ Rev.\ D {\bf 70}, 113009 (2004)
  [arXiv:hep-ph/0406323].
%
\bibitem{Castanier:2001sf}
  C.~Castanier, P.~Gay, P.~Lutz and J.~Orloff,
  arXiv:hep-ex/0101028.
%
\bibitem{Baur:2003gp}
  U.~Baur, T.~Plehn and D.~L.~Rainwater,
  Phys.\ Rev.\ D {\bf 69}, 053004 (2004)
  [arXiv:hep-ph/0310056].
%
\bibitem{ilc_calc}
  F.~Boudjema {\it et al.},
  Phys.\ Lett.\ B {\bf 600}, 65 (2004)
  [arXiv:hep-ph/0407065];
  G.~Belanger {\it et al.},
  Phys.\ Lett.\ B {\bf 576}, 152 (2003)
  [arXiv:hep-ph/0309010];
  G.~Belanger, F.~Boudjema, J.~Fujimoto, T.~Ishikawa, T.~Kaneko, K.~Kato
and Y.~Shimizu, 
  Phys.\ Lett.\ B {\bf 559}, 252 (2003)
  [arXiv:hep-ph/0212261];
  A.~Denner, J.~K\"ublbeck, R.~Mertig and M.~B\"ohm,
  Z.\ Phys.\ C {\bf 56}, 261 (1992);
  B.~A.~Kniehl,
  Z.\ Phys.\ C {\bf 55}, 605 (1992);
  A.~Denner, S.~Dittmaier, M.~Roth and M.~M.~Weber,
  Nucl.\ Phys.\ B {\bf 660}, 289 (2003)
  [arXiv:hep-ph/0302198];
  A.~Denner, S.~Dittmaier, M.~Roth and M.~M.~Weber,
  Phys.\ Lett.\ B {\bf 560}, 196 (2003)
  [arXiv:hep-ph/0301189];
  R.~Y.~Zhang, W.~G.~Ma, H.~Chen, Y.~B.~Sun and H.~S.~Hou,
  Phys.\ Lett.\ B {\bf 578}, 349 (2004)
  [arXiv:hep-ph/0308203];
  F.~Boudjema {\it et al.},
{\it In the Proceedings of 2005 International Linear Collider Workshop (LCWS 2005), Stanford, California, 18-22 Mar 2005, pp 0601}
  [arXiv:hep-ph/0510184];
  S.~Dawson and L.~Reina,
  Phys.\ Rev.\ D {\bf 59}, 054012 (1999)
  [arXiv:hep-ph/9808443];
  Y.~You, W.~G.~Ma, H.~Chen, R.~Y.~Zhang, S.~Yan-Bin and H.~S.~Hou,
  Phys.\ Lett.\ B {\bf 571}, 85 (2003)
  [arXiv:hep-ph/0306036];
  G.~Belanger {\it et al.},
  Phys.\ Lett.\ B {\bf 571}, 163 (2003)
  [arXiv:hep-ph/0307029];
  A.~Denner, S.~Dittmaier, M.~Roth and M.~M.~Weber,
  Phys.\ Lett.\ B {\bf 575}, 290 (2003)
  [arXiv:hep-ph/0307193].
%
\bibitem{Barate:2003sz}
  R.~Barate {\it et al.}  [LEP Working Group for Higgs boson searches],
  Phys.\ Lett.\ B {\bf 565}, 61 (2003)
  [arXiv:hep-ex/0306033].
%
\bibitem{Awramik:2006uz}
  M.~Awramik, M.~Czakon and A.~Freitas,
  JHEP {\bf 0611}, 048 (2006)
  [arXiv:hep-ph/0608099] and
  Phys.\ Lett.\ B {\bf 642}, 563 (2006)
  [arXiv:hep-ph/0605339];
  W.~Hollik, U.~Meier and S.~Uccirati,
  Nucl.\ Phys.\ B {\bf 731}, 213 (2005)
  [arXiv:hep-ph/0507158] and
  arXiv:hep-ph/0610312.
%
\bibitem{Group:2006mx}
  J.~Alcaraz {\it et al.} [The LEP Electroweak Working Group],
  arXiv:hep-ex/0612034.
%
\bibitem{Maltoni:2002qb}
  F.~Maltoni and T.~Stelzer,
  JHEP {\bf 0302}, 027 (2003)
  [arXiv:hep-ph/0208156].
%
\bibitem{Yuasa:1999rg}
  F.~Yuasa {\it et al.},
  Prog.\ Theor.\ Phys.\ Suppl.\  {\bf 138}, 18 (2000)
  [arXiv:hep-ph/0007053].
%
\bibitem{Pukhov:2004ca}
  A.~Pukhov,
  arXiv:hep-ph/0412191.
%
\bibitem{Boos:2004kh}
  E.~Boos {\it et al.}  [CompHEP Collaboration],
  Nucl.\ Instrum.\ Meth.\ A {\bf 534}, 250 (2004)
  [arXiv:hep-ph/0403113];
  A.~Pukhov {\it et al.},
  arXiv:hep-ph/9908288.
%
\bibitem{kilian}
W.~Kilian, LC-TOOL-2001-039.
%
\bibitem{Mangano:2002ea}
  M.~L.~Mangano, M.~Moretti, F.~Piccinini, R.~Pittau and A.~D.~Polosa,
  JHEP {\bf 0307}, 001 (2003)
  [arXiv:hep-ph/0206293].
%
\bibitem{Kryukov:2004md}
  A.~Kryukov and L.~Shamardin,
  Nucl.\ Instrum.\ Meth.\ A {\bf 534}, 329 (2004).
%
\bibitem{Belanger:2003sd}
  G.~Belanger, F.~Boudjema, J.~Fujimoto, T.~Ishikawa, T.~Kaneko, K.~Kato and Y.~Shimizu,
  Phys.\ Rept.\  {\bf 430}, 117 (2006)
  [arXiv:hep-ph/0308080].
%
\bibitem{Hahn:2000kx}
  T.~Hahn,
  Comput.\ Phys.\ Commun.\  {\bf 140}, 418 (2001)
  [arXiv:hep-ph/0012260].
%
\bibitem{Mertig:1990an}
  R.~Mertig, M.~B\"ohm and A.~Denner,
  Comput.\ Phys.\ Commun.\  {\bf 64}, 345 (1991).
%
\bibitem{Hahn:1998yk}
  T.~Hahn and M.~Perez-Victoria,
  Comput.\ Phys.\ Commun.\  {\bf 118}, 153 (1999)
  [arXiv:hep-ph/9807565].
%
\bibitem{Tentyukov:2003bx}
  M.~Tentyukov and J.~Fleischer,
  Comput.\ Phys.\ Commun.\  {\bf 160}, 167 (2004)
  [arXiv:hep-ph/0311111].
%
\bibitem{Vermaseren:2000nd}
  J.~A.~M.~Vermaseren,
  arXiv:math-ph/0010025.
%
\bibitem{kurihara}
K.~Kato, talk given at ``LoopFest III'', KITP, Santa Barbara, April~2004.
%
\bibitem{Kurihara:2006kt}
  Y.~Kurihara {\it et al.},
  Nucl.\ Phys.\ Proc.\ Suppl.\  {\bf 157}, 231 (2006).
%
\bibitem{walter}
W.~Giele, private communication
%
\bibitem{stefan}
S.~Dittmaier, talk given at ``LoopFest V'', SLAC, June~2006.
%
\bibitem{twojet}
  S.~D.~Ellis, Z.~Kunszt and D.~E.~Soper,
  Phys.\ Rev.\ Lett.\  {\bf 69}, 1496 (1992).
%
\bibitem{threejet}
  W.~B.~Kilgore and W.~T.~Giele,
  Phys.\ Rev.\ D {\bf 55}, 7183 (1997)
  [arXiv:hep-ph/9610433];
  Z.~Nagy,
  Phys.\ Rev.\ D {\bf 68}, 094002 (2003)
  [arXiv:hep-ph/0307268].
%
\bibitem{Ellis:2006ss}
  R.~K.~Ellis, W.~T.~Giele and G.~Zanderighi,
  JHEP {\bf 0605}, 027 (2006)
  [arXiv:hep-ph/0602185].
%
\bibitem{Ellis:2005zh}
  R.~K.~Ellis, W.~T.~Giele and G.~Zanderighi,
  Phys.\ Rev.\ D {\bf 73}, 014027 (2006)
  [arXiv:hep-ph/0508308].
%
\bibitem{Jager:2006zc}
  B.~J\"ager, C.~Oleari and D.~Zeppenfeld,
  JHEP {\bf 0607}, 015 (2006)
  [arXiv:hep-ph/0603177].
%
\bibitem{Kauer:2000hi}
  N.~Kauer, T.~Plehn, D.~L.~Rainwater and D.~Zeppenfeld,
  Phys.\ Lett.\ B {\bf 503}, 113 (2001)
  [arXiv:hep-ph/0012351].
%
\bibitem{Ciafaloni:1998xg}
  P.~Ciafaloni and D.~Comelli,
  Phys.\ Lett.\ B {\bf 446}, 278 (1999)
  [arXiv:hep-ph/9809321].
%
\bibitem{Ciafaloni:2000df}
  M.~Ciafaloni, P.~Ciafaloni and D.~Comelli,
  Phys.\ Rev.\ Lett.\  {\bf 84}, 4810 (2000)
  [arXiv:hep-ph/0001142];
  Nucl.\ Phys.\ B {\bf 589}, 359 (2000)
  [arXiv:hep-ph/0004071].
%
\bibitem{Ciafaloni:2000gm}
  M.~Ciafaloni, P.~Ciafaloni and D.~Comelli,
  Phys.\ Lett.\ B {\bf 501}, 216 (2001)
  [arXiv:hep-ph/0007096].
%
\bibitem{Denner:2000jv}
  A.~Denner and S.~Pozzorini,
  Eur.\ Phys.\ J.\ C {\bf 18}, 461 (2001)
  [arXiv:hep-ph/0010201];
  Eur.\ Phys.\ J.\ C {\bf 21}, 63 (2001)
  [arXiv:hep-ph/0104127].
%
\bibitem{Kuhn:1999de}
  J.~H.~K\"uhn and A.~A.~Penin,
  arXiv:hep-ph/9906545;
  J.~H.~K\"uhn, A.~A.~Penin and V.~A.~Smirnov,
  Eur.\ Phys.\ J.\ C {\bf 17}, 97 (2000)
  [arXiv:hep-ph/9912503];
  J.~H.~K\"uhn, S.~Moch, A.~A.~Penin and V.~A.~Smirnov,
  Nucl.\ Phys.\ B {\bf 616}, 286 (2001)
  [Erratum-ibid.\ B {\bf 648}, 455 (2003)]
  [arXiv:hep-ph/0106298];
  B.~Jantzen, J.~H.~K\"uhn, A.~A.~Penin and V.~A.~Smirnov,
  Phys.\ Rev.\ D {\bf 72}, 051301 (2005)
  [Erratum-ibid.\ D {\bf 74}, 019901 (2006)]
  [arXiv:hep-ph/0504111], and
  Nucl.\ Phys.\ B {\bf 731}, 188 (2005)
  [Erratum-ibid.\ B {\bf 752}, 327 (2006)]
  [arXiv:hep-ph/0509157].
%
\bibitem{Melles:2001ye}
  For a review, see M.~Melles,
  Phys.\ Rept.\  {\bf 375}, 219 (2003)
  [arXiv:hep-ph/0104232].
%
\bibitem{Moretti:2005ut}
  S.~Moretti, M.~R.~Nolten and D.~A.~Ross,
  Phys.\ Rev.\ D {\bf 74}, 097301 (2006)
  [arXiv:hep-ph/0503152];
  Nucl.\ Phys.\ B {\bf 759}, 50 (2006)
  [arXiv:hep-ph/0606201].
%
\bibitem{Kuhn:2005gv}
  J.~H.~K\"uhn, A.~Kulesza, S.~Pozzorini and M.~Schulze,
  JHEP {\bf 0603}, 059 (2006)
  [arXiv:hep-ph/0508253].
%
\bibitem{Maina:2004rb}
  E.~Maina, S.~Moretti and D.~A.~Ross,
  Phys.\ Lett.\ B {\bf 593}, 143 (2004)
  [Erratum-ibid.\ B {\bf 614}, 216 (2005)]
  [arXiv:hep-ph/0403050].
%
\bibitem{Kuhn:2004em}
  J.~H.~K\"uhn, A.~Kulesza, S.~Pozzorini and M.~Schulze,
  Phys.\ Lett.\ B {\bf 609}, 277 (2005)
  [arXiv:hep-ph/0408308];
  Nucl.\ Phys.\ B {\bf 727}, 368 (2005)
  [arXiv:hep-ph/0507178].
%
\bibitem{Baur:2001ze}
  U.~Baur, O.~Brein, W.~Hollik, C.~Schappacher and D.~Wackeroth,
  Phys.\ Rev.\ D {\bf 65}, 033007 (2002)
  [arXiv:hep-ph/0108274].
%
\bibitem{kramer}
 S.~Dittmaier and M.~Kr\"amer,
  Phys.\ Rev.\ D {\bf 65}, 073007 (2002)
  [arXiv:hep-ph/0109062].
%
\bibitem{Baur:2004ig}
  U.~Baur and D.~Wackeroth,
  Phys.\ Rev.\ D {\bf 70}, 073015 (2004)
  [arXiv:hep-ph/0405191].
%
\bibitem{Arbuzov:2005dd}
  A.~Arbuzov, D.~Bardin, S.~Bondarenko, P.~Christova, L.~Kalinovskaya,
G.~Nanava and R.~Sadykov, 
  Eur.\ Phys.\ J.\ C {\bf 46}, 407 (2006)
  [arXiv:hep-ph/0506110].
%
\bibitem{Zykunov:2005tc}
  V.~A.~Zykunov,
  arXiv:hep-ph/0509315 and
  Phys.\ Atom.\ Nucl.\  {\bf 69}, 1522 (2006).
%
\bibitem{CarloniCalame:2006zq}
  C.~M.~Carloni Calame, G.~Montagna, O.~Nicrosini and A.~Vicini,
  arXiv:hep-ph/0609170.
%
\bibitem{Accomando:2001fn}
  E.~Accomando, A.~Denner and S.~Pozzorini,
  Phys.\ Rev.\ D {\bf 65}, 073003 (2002)
  [arXiv:hep-ph/0110114];
  W.~Hollik and C.~Meier,
  Phys.\ Lett.\ B {\bf 590}, 69 (2004)
  [arXiv:hep-ph/0402281].
%
\bibitem{wgam}
  E.~Accomando, A.~Denner and C.~Meier,
  Eur.\ Phys.\ J.\ C {\bf 47}, 125 (2006) [arXiv:hep-ph/0509234].
%
\bibitem{Accomando:2004de}
  E.~Accomando, A.~Denner and A.~Kaiser,
  Nucl.\ Phys.\ B {\bf 706}, 325 (2005)
  [arXiv:hep-ph/0409247].
%
\bibitem{Kuhn:2005it}
  J.~H.~K\"uhn, A.~Scharf and P.~Uwer,
  Eur.\ Phys.\ J.\ C {\bf 45}, 139 (2006)
  [arXiv:hep-ph/0508092];
  W.~Bernreuther, M.~F\"ucker and Z.~G.~Si,
  Phys.\ Lett.\ B {\bf 633}, 54 (2006)
  [arXiv:hep-ph/0508091].
\bibitem{mnr}
  S.~Moretti, M.~R.~Nolten and D.~A.~Ross,
  Phys.\ Lett.\ B {\bf 639}, 513 (2006)
  [arXiv:hep-ph/0603083].
\bibitem{Beenakker:1993yr}
  W.~Beenakker, A.~Denner, W.~Hollik, R.~Mertig, T.~Sack and D.~Wackeroth,
  Nucl.\ Phys.\ B {\bf 411}, 343 (1994).
%
\bibitem{Bernreuther:2006vg}
  W.~Bernreuther, M.~F\"ucker and Z.~G.~Si,
  arXiv:hep-ph/0610334;
  J.~H.~K\"uhn, A.~Scharf and P.~Uwer,
  arXiv:hep-ph/0610335.
%
\bibitem{Beccaria:2006ir}
  M.~Beccaria, G.~Macorini, F.~M.~Renard and C.~Verzegnassi,
  Phys.\ Rev.\ D {\bf 74}, 013008 (2006)
  [arXiv:hep-ph/0605108];
  arXiv:hep-ph/0609189.
%
\bibitem{Beccaria:2006dt}
  M.~Beccaria, G.~Macorini, F.~M.~Renard and C.~Verzegnassi,
  Phys.\ Rev.\ D {\bf 73}, 093001 (2006)
  [arXiv:hep-ph/0601175].
%
\bibitem{Ciafaloni:2006qu}
  P.~Ciafaloni and D.~Comelli,
  JHEP {\bf 0609}, 055 (2006)
  [arXiv:hep-ph/0604070].
%
\bibitem{Baur:2006sn}
  U.~Baur,
  Phys.\ Rev.\ D {\bf 75}, 013005 (2007)
  [arXiv:hep-ph/0611241].
%
\bibitem{Hawkings:1999ac}
  R.~Hawkings and K.~M\"onig,
  Eur.\ Phys.\ J.\ directC {\bf 1}, 8 (1999)
  [arXiv:hep-ex/9910022].
%
\bibitem{Monig:2003ze}
  K.~M\"onig,
  arXiv:hep-ph/0303023.
%
\bibitem{Stirling:1995xp}
  W.~J.~Stirling,
  Nucl.\ Phys.\ B {\bf 456}, 3 (1995)
  [arXiv:hep-ph/9503320].
%
\bibitem{Grunewald:2000ju}
  M.~W.~Gr\"unewald {\it et al.},
  arXiv:hep-ph/0005309.
%
\bibitem{Denner:2005es}
  A.~Denner, S.~Dittmaier, M.~Roth and L.~H.~Wieders,
  Phys.\ Lett.\ B {\bf 612}, 223 (2005)
  [arXiv:hep-ph/0502063] and
  Nucl.\ Phys.\ B {\bf 724}, 247 (2005)
  [arXiv:hep-ph/0505042].
%
\bibitem{Denner:2005nn}
  A.~Denner and S.~Dittmaier,
  Nucl.\ Phys.\ B {\bf 734}, 62 (2006)
  [arXiv:hep-ph/0509141].
%
\bibitem{Fadin:1995fp}
  V.~S.~Fadin, V.~A.~Khoze, A.~D.~Martin and W.~J.~Stirling,
  Phys.\ Lett.\ B {\bf 363}, 112 (1995)
  [arXiv:hep-ph/9507422].
%
\bibitem{Bardin:1993mc}
  D.~Y.~Bardin, W.~Beenakker and A.~Denner,
  Phys.\ Lett.\ B {\bf 317}, 213 (1993).
%
\bibitem{Anastasiou:2000mf}
  K.~G.~Chetyrkin and F.~V.~Tkachov,
  Nucl.\ Phys.\ B {\bf 192} (1981) 159;
  K.~G.~Chetyrkin, A.~L.~Kataev and F.~V.~Tkachov,
  Nucl.\ Phys.\ B {\bf 174}, 345 (1980);
  C.~Anastasiou, T.~Gehrmann, C.~Oleari, E.~Remiddi and J.~B.~Tausk,
  Nucl.\ Phys.\ B {\bf 580}, 577 (2000)
  [arXiv:hep-ph/0003261];
  T.~Gehrmann and E.~Remiddi,
  Nucl.\ Phys.\ B {\bf 580}, 485 (2000)
  [arXiv:hep-ph/9912329];
  V.~A.~Smirnov and O.~L.~Veretin,
  Nucl.\ Phys.\ B {\bf 566}, 469 (2000)
  [arXiv:hep-ph/9907385];
  T.~G.~Birthwright, E.~W.~N.~Glover and P.~Marquard,
  JHEP {\bf 0409}, 042 (2004)
  [arXiv:hep-ph/0407343];
  M.~Caffo, H.~Czyz, S.~Laporta and E.~Remiddi,
  Nuovo Cim.\ A {\bf 111}, 365 (1998)
  [arXiv:hep-th/9805118];
  S.~Laporta,
  Phys.\ Lett.\ B {\bf 504}, 188 (2001)
  [arXiv:hep-ph/0102032];
  C.~Anastasiou and A.~Lazopoulos,
  JHEP {\bf 0407}, 046 (2004)
  [arXiv:hep-ph/0404258];
  V.~A.~Smirnov,
  Phys.\ Lett.\ B {\bf 460}, 397 (1999)
  [arXiv:hep-ph/9905323], 
  Phys.\ Lett.\ B {\bf 491}, 130 (2000)
  [arXiv:hep-ph/0007032] and
  Phys.\ Lett.\ B {\bf 500}, 330 (2001)
  [arXiv:hep-ph/0011056];
  J.~B.~Tausk,
  Phys.\ Lett.\ B {\bf 469}, 225 (1999)
  [arXiv:hep-ph/9909506];
  C.~Anastasiou, E.~W.~N.~Glover and C.~Oleari,
  Nucl.\ Phys.\ B {\bf 575}, 416 (2000)
  [Erratum-ibid.\ B {\bf 585}, 763 (2000)]
  [arXiv:hep-ph/9912251];
  T.~Binoth and G.~Heinrich,
  Nucl.\ Phys.\ B {\bf 680}, 375 (2004)
  [arXiv:hep-ph/0305234] and
  Nucl.\ Phys.\ B {\bf 585}, 741 (2000)
  [arXiv:hep-ph/0004013].
%
\bibitem{glover}
  C.~Anastasiou, E.~W.~N.~Glover, C.~Oleari and M.~E.~Tejeda-Yeomans,
  Nucl.\ Phys.\ B {\bf 601}, 318 (2001)
  [arXiv:hep-ph/0010212];
  Nucl.\ Phys.\ B {\bf 601}, 341 (2001)
  [arXiv:hep-ph/0011094] and
  Nucl.\ Phys.\ B {\bf 605}, 486 (2001)
  [arXiv:hep-ph/0101304];
  E.~W.~N.~Glover, C.~Oleari and M.~E.~Tejeda-Yeomans,
  Nucl.\ Phys.\ B {\bf 605}, 467 (2001)
  [arXiv:hep-ph/0102201];
  C.~Anastasiou, E.~W.~N.~Glover and M.~E.~Tejeda-Yeomans,
  Nucl.\ Phys.\ B {\bf 629}, 255 (2002)
  [arXiv:hep-ph/0201274].
\bibitem{Garland:2002ak}
  L.~W.~Garland, T.~Gehrmann, E.~W.~N.~Glover, A.~Koukoutsakis and E.~Remiddi,
  Nucl.\ Phys.\ B {\bf 642}, 227 (2002)
  [arXiv:hep-ph/0206067];
  S.~Moch, P.~Uwer and S.~Weinzierl,
  Phys.\ Rev.\ D {\bf 66}, 114001 (2002)
  [arXiv:hep-ph/0207043].
%
\bibitem{Bern:2000ie}
  Z.~Bern, L.~J.~Dixon and A.~Ghinculov,
  Phys.\ Rev.\ D {\bf 63}, 053007 (2001)
  [arXiv:hep-ph/0010075].
%
\bibitem{Gehrmann-DeRidder:2005cm}
  A.~Gehrmann-De Ridder, T.~Gehrmann and E.~W.~N.~Glover,
  JHEP {\bf 0509}, 056 (2005)
  [arXiv:hep-ph/0505111].
%
\bibitem{Anastasiou:2003gr}
  C.~Anastasiou, K.~Melnikov and F.~Petriello,
  Phys.\ Rev.\ D {\bf 69}, 076010 (2004)
  [arXiv:hep-ph/0311311].
%
\bibitem{Binoth:2004jv}
  T.~Binoth and G.~Heinrich,
  Nucl.\ Phys.\ B {\bf 693}, 134 (2004)
  [arXiv:hep-ph/0402265].
%
\bibitem{Anastasiou:2005qj}
  C.~Anastasiou, K.~Melnikov and F.~Petriello,
  Nucl.\ Phys.\ B {\bf 724}, 197 (2005)
  [arXiv:hep-ph/0501130].
%
\bibitem{Dittmar:1997md}
  M.~Dittmar, F.~Pauss and D.~Z\"urcher,
  Phys.\ Rev.\ D {\bf 56}, 7284 (1997)
  [arXiv:hep-ex/9705004];
  V.~A.~Khoze, A.~D.~Martin, R.~Orava and M.~G.~Ryskin,
  Eur.\ Phys.\ J.\ C {\bf 19}, 313 (2001)
  [arXiv:hep-ph/0010163];
  W.~T.~Giele and S.~A.~Keller,
  arXiv:hep-ph/0104053.
%
\bibitem{Buscher:2005re}
  V.~B\"uscher and K.~Jakobs,
  Int.\ J.\ Mod.\ Phys.\ A {\bf 20}, 2523 (2005)
  [arXiv:hep-ph/0504099].
%
%
\bibitem{tree}
F.~Cachazo, P.~Svr\v{c}ek and E.~Witten,
JHEP {\bf 0409}, 006 (2004)
[arXiv:hep-th/0403047];
%
C.~J.\ Zhu,
JHEP {\bf 0404}, 032 (2004)
[arXiv:hep-th/0403115];
%
G.\ Georgiou and V.\ V.\ Khoze,
JHEP {\bf 0405}, 070 (2004)
[arXiv:hep-th/0404072];
%
J.~B.~Wu and C.~J.~Zhu,
JHEP {\bf 0407}, 032 (2004)
[arXiv:hep-th/0406085];
%
J.~B.~Wu and C.~J.~Zhu,
JHEP {\bf 0409}, 063 (2004)
[arXiv:hep-th/0406146];
%
D.\ A.\ Kosower,
Phys.\ Rev.\ D {\bf 71}, 045007 (2005)
[arXiv:hep-th/0406175];
%
G.~Georgiou, E.~W.~N.~Glover and V.~V.~Khoze,
JHEP {\bf 0407}, 048 (2004)
[arXiv:hep-th/0407027]; 
%
Y.~Abe, V.~P.~Nair and M.~I.~Park,
Phys.\ Rev.\ D {\bf 71}, 025002 (2005)
[arXiv:hep-th/0408191];
L.~J.~Dixon, E.~W.~N.~Glover and V.~V.~Khoze,
JHEP {\bf 0412}, 015 (2004)
[arXiv:hep-th/0411092];
%
S.~D.~Badger, E.~W.~N.~Glover and V.~V.~Khoze,
JHEP {\bf 0503}, 023 (2005)
[arXiv:hep-th/0412275];
Z.~Bern, D.~Forde, D.~A.~Kosower and P.~Mastrolia,
Phys.\ Rev.\ D {\bf 72}, 025006 (2005)
  [arXiv:hep-ph/0412167];
R.~Roiban, M.~Spradlin and A.~Volovich,
Phys.\ Rev.\ Lett.\ {\bf 94}, 102002 (2005)
[arXiv:hep-th/0412265];
R.~Britto, F.~Cachazo and B.~Feng,
Nucl.\ Phys.\ B {\bf 715}, 499 (2005)
  [arXiv:hep-th/0412308];
R.~Britto, F.~Cachazo, B.~Feng and E.~Witten,
Phys.\ Rev.\ Lett.\  {\bf 94}, 181602 (2005)
  [arXiv:hep-th/0501052];
M.~Luo and C.~Wen,
JHEP {\bf 0503}, 004 (2005)
[arXiv:hep-th/0501121];
Phys.\ Rev.\ D {\bf 71}, 091501 (2005)
  [arXiv:hep-th/0502009];
R.~Britto, B.~Feng, R.~Roiban, M.~Spradlin and A.~Volovich,
Phys.\ Rev.\ D {\bf 71}, 105017 (2005)
  [arXiv:hep-th/0503198];
S.~D.~Badger, E.~W.~N.~Glover, V.~V.~Khoze and P.~Svr\v{c}ek,
JHEP {\bf 0507}, 025 (2005)
  [arXiv:hep-th/0504159].
%
\bibitem{oneloop}
A.~Brandhuber, B.~Spence and G.~Travaglini,
Nucl.\ Phys.\ B {\bf 706}, 150 (2005)
[arXiv:hep-th/0407214];
R.~Britto, F.~Cachazo and B.~Feng,
Phys.\ Rev.\ D {\bf 71}, 025012 (2005)
[arXiv:hep-th/0410179];
Z.~Bern, V.~Del Duca, L.~J.~Dixon and D.~A.~Kosower,
Phys.\ Rev.\ D {\bf 71}, 045006 (2005)
[arXiv:hep-th/0410224];
R.~Britto, F.~Cachazo and B.~Feng,
  Nucl.\ Phys.\ B {\bf 725}, 275 (2005)
  [arXiv:hep-th/0412103];
Z.~Bern, L.~J.~Dixon and D.~A.~Kosower,
Phys.\ Rev.\ D {\bf 72}, 045014 (2005)
  [arXiv:hep-th/0412210];
  Phys.\ Rev.\ D {\bf 73}, 065013 (2006)
  [arXiv:hep-ph/0507005]; 
C.~Quigley and M.~Rozali,
JHEP {\bf 0501}, 053 (2005)
[arXiv:hep-th/0410278];
%
J.~Bedford, A.~Brandhuber, B.~Spence and G.~Travaglini,
Nucl.\ Phys.\ B {\bf 706}, 100 (2005)
[arXiv:hep-th/0410280];
Nucl.\ Phys.\ B {\bf 712}, 59 (2005)
[arXiv:hep-th/0412108];
S.~J.~Bidder, N.~E.~J.~Bjerrum-Bohr, L.~J.~Dixon and D.~C.~Dunbar,
Phys.\ Lett.\ B {\bf 606}, 189 (2005)
[arXiv:hep-th/0410296];
%
S.~J.~Bidder, N.~E.~J.~Bjerrum-Bohr, D.~C.~Dunbar and W.~B.~Perkins,
Phys.\ Lett.\ B {\bf 608}, 151 (2005)
[arXiv:hep-th/0412023];
Phys.\ Lett.\ B {\bf 612}, 75 (2005)
[arXiv:hep-th/0502028];
R.~Britto, E.~Buchbinder, F.~Cachazo and B.~Feng,
 Phys.\ Rev.\ D {\bf 72}, 065012 (2005)
  [arXiv:hep-ph/0503132];
  C.~F.~Berger, Z.~Bern, L.~J.~Dixon, D.~Forde and D.~A.~Kosower,
  Phys.\ Rev.\ D {\bf 74}, 036009 (2006)
  [arXiv:hep-ph/0604195].
  C.~F.~Berger, Z.~Bern, L.~J.~Dixon, D.~Forde and D.~A.~Kosower,
  arXiv:hep-ph/0607014.
%
\bibitem{Witten:2003nn}
  E.~Witten,
  Commun.\ Math.\ Phys.\  {\bf 252}, 189 (2004)
  [arXiv:hep-th/0312171].
%
\bibitem{dr1lA} R.~Barbieri and L. Maiani,  
                Nucl. Phys. B {\bf 224} (1983) 32; 
                C.~Lim, T.~Inami and N.~Sakai, 
                Phys. Rev. D {\bf 29} (1984) 1488; 
                E.~Eliasson, 
                 Phys. Lett. B {\bf 147} (1984) 65; 
                Z.~Hioki, 
                Prog. Theo. Phys. {\bf 73} (1985) 1283; 
                J.~Grifols and J.~Sol\`a, 
                Nucl. Phys. B {\bf  253} (1985) 47; 
                B.~Lynn, M.~Peskin and R.~Stuart, 
                CERN Report 86-02, p. 90; 
                R.~Barbieri, M.~Frigeni, F.~Giuliani and H.~Haber, 
                Nucl. Phys. B {\bf 341} (1990) 309;   
                M.~Drees and K.~Hagiwara, 
                Phys. Rev. D {\bf 42} (1990) 1709.
%
\bibitem{dr1lB} M.~Drees, K.~Hagiwara and A.~Yamada, 
                Phys. Rev. D {\bf 45} (1992) 1725.
%
\bibitem{MWMSSM1LA} P.~Chankowski, A.~Dabelstein, W.~Hollik,
                    W.~M\"osle, S.~Pokorski and  J.~Rosiek, 
                    Nucl. Phys. B {\bf 417} (1994) 101.
%
\bibitem{MWMSSM1LB} D.~Garcia and J.~Sola, 
                     Mod. Phys. Lett. A {\bf 9} (1994) 211.
%
\bibitem{pierce} D.~Pierce, J.~Bagger, K.~Matchev and R.~Zhang,
                  Nucl. Phys. B {\bf  491} (1997) 3,
                 hep-ph/9606211.
%
\bibitem{Heinemeyer:2002jq}
  S.~Heinemeyer and G.~Weiglein,
  JHEP {\bf 0210}, 072 (2002)
  [arXiv:hep-ph/0209305].
%
\bibitem{Frank:2006yh}
  M.~Frank, T.~Hahn, S.~Heinemeyer, W.~Hollik, H.~Rzehak and G.~Weiglein,
  arXiv:hep-ph/0611326.
%
\bibitem{Heinemeyer:2004gx}
  S.~Heinemeyer, W.~Hollik and G.~Weiglein,
  Phys.\ Rept.\  {\bf 425}, 265 (2006)
  [arXiv:hep-ph/0412214].
%
\bibitem{Diaz:2000hi}
  M.~A.~Diaz, S.~F.~King and D.~A.~Ross,
  Phys.\ Rev.\ D {\bf 64}, 017701 (2001)
  [arXiv:hep-ph/0008117];
  T.~Blank and W.~Hollik,
  arXiv:hep-ph/0011092.
%
\bibitem{Oller:2005xg}
  W.~\"Oller, H.~Eberl and W.~Majerotto,
  Phys.\ Rev.\ D {\bf 71}, 115002 (2005)
  [arXiv:hep-ph/0504109].
%
\bibitem{Beenakker:1996ch}
  W.~Beenakker, R.~H\"opker, M.~Spira and P.~M.~Zerwas,
  Nucl.\ Phys.\ B {\bf 492}, 51 (1997)
  [arXiv:hep-ph/9610490];
  W.~Beenakker, M.~Kr\"amer, T.~Plehn, M.~Spira and P.~M.~Zerwas,
  Nucl.\ Phys.\ B {\bf 515}, 3 (1998)
  [arXiv:hep-ph/9710451];
  W.~Beenakker, M.~Klasen, M.~Kr\"amer, T.~Plehn, M.~Spira and P.~M.~Zerwas,
  Phys.\ Rev.\ Lett.\  {\bf 83}, 3780 (1999)
  [arXiv:hep-ph/9906298];
  W.~Beenakker, M.~Klasen, M.~Kr\"amer, T.~Plehn, M.~Spira and P.~M.~Zerwas,
  Phys.\ Rev.\ Lett.\  {\bf 83}, 3780 (1999)
  [arXiv:hep-ph/9906298];
  M.~Spira,
  arXiv:hep-ph/0211145;
  T.~Plehn,
  Czech.\ J.\ Phys.\  {\bf 55}, B213 (2005)
  [arXiv:hep-ph/0410063].
%
\bibitem{Muhlleitner:2003vg}
  M.~M\"uhlleitner, A.~Djouadi and Y.~Mambrini,
  Comput.\ Phys.\ Commun.\  {\bf 168}, 46 (2005)
  [arXiv:hep-ph/0311167].
%
\bibitem{Cho:2006sx}
  G.~C.~Cho, K.~Hagiwara, J.~Kanzaki, T.~Plehn, D.~Rainwater and T.~Stelzer,
  Phys.\ Rev.\ D {\bf 73}, 054002 (2006)
  [arXiv:hep-ph/0601063].
%
\bibitem{Baur:2002rb}
  U.~Baur, T.~Plehn and D.~L.~Rainwater,
  Phys.\ Rev.\ Lett.\  {\bf 89}, 151801 (2002)
  [arXiv:hep-ph/0206024] and
  Phys.\ Rev.\ D {\bf 67}, 033003 (2003)
  [arXiv:hep-ph/0211224].
%
\end{thebibliography}
\end{document}